\pdfoutput=1

\documentclass{ws-ijbc}
\usepackage{graphicx}
\usepackage{multicol}
\usepackage{hyperref}

\hypersetup{pdftitle = The title of my PDF, pdfauthor = My name, pdfsubject= The subject, pdfkeywords = keyword1 keyword2 keyword3}

\hypersetup{colorlinks = true, linkcolor = blue, anchorcolor = red, citecolor = blue, filecolor = red, pagecolor = red, urlcolor = blue}

\begin{document}

\catchline{}{}{}{}{} 

\markboth{E.E. Zotos}{Basins of convergence of equilibrium points in the generalized Hill problem}

\title{Basins of Convergence of Equilibrium Points \\
in the Generalized Hill Problem}

\author{Euaggelos E. Zotos}

\address{Department of Physics, School of Science, \\
Aristotle University of Thessaloniki, \\
GR-541 24, Thessaloniki, Greece \\
Corresponding author's email: {evzotos@physics.auth.gr}}

\maketitle

\begin{history}
\received{Received August 19, 2017; Revised October 10, 2017}
\end{history}

\begin{abstract}
The Newton-Raphson basins of attraction, associated with the libration points (attractors), are revealed in the generalized Hill problem. The parametric variation of the position and the linear stability of the equilibrium points is determined, when the value of the perturbation parameter $\epsilon$ varies. The multivariate Newton-Raphson iterative scheme is used to determine the attracting domains on several types of two-dimensional planes. A systematic and thorough numerical investigation is performed in order to demonstrate the influence of the perturbation parameter on the geometry as well as of the basin entropy of the basins of convergence. The correlations between the basins of attraction and the corresponding required number of iterations are also illustrated and discussed. Our numerical analysis strongly indicates that the evolution of the attracting regions in this dynamical system is an extremely complicated yet very interesting issue.
\end{abstract}


\twocolumn

\section{Introduction}
\label{intro}

Undoubtedly, one of the most intriguing as well as important fields in dynamical astronomy and celestial mechanics is the few-body problem and especially the version of the circular restricted three-body problem \cite{S67}. This is true if we take into account that this problem has numerous applications in many research fields, such as molecular physics, chaos theory, planetary physics, or even stellar and galactic dynamics. This is exactly why this topic remains active and stimulating even today.

The Hill limiting case is in fact a simplified modification of the three-body problem which focus on the vicinity of the secondary (e.g., \cite{H86,PH86,PH87}). This allows us to study the motion of the test particles in the neighborhood of the equilibrium points $L_1$ and $L_2$. At this point, it should be emphasized that the Hill approximation is valid only when the mass of the secondary is much smaller compared with the mass of the primary body $(m_2 \ll m_1)$. One can directly obtain the Hill model from the classical three-body problem by translating the origin to the center of the secondary and also by rescaling the coordinates by a factor $\mu^{1/3}$, where $\mu = m_2/(m_1 + m_2)$ is the mass ratio.

Knowing the basins of convergence, associated with the libration points, is an issue of great importance, since the attracting domains reflect some of the most intrinsic properties of the dynamical system. For obtaining the basins of attraction one should use an iterative scheme (e.g., the Newton-Raphson method) and scan a set of initial conditions in order to reveal their final states (attractors). Over the past years a large number of studies have been devoted on determining the Newton-Raphson basins of convergence in many types of dynamical systems, such as the Hill's problem \cite{D10}, the Sitnikov problem \cite{DKMP12}, the restricted three-body problem with oblateness and radiation pressure \cite{Z16}, the electromagnetic Copenhagen problem \cite{KGK12,Z17b}, the photogravitational Copenhagen problem \cite{K08}, the four-body problem \cite{BP11,KK14,Z17a}, the photogravitational four-body problem \cite{APHS16}, the ring problem of $N + 1$ bodies \cite{CK07,GKK09}, or even the restricted 2+2 body problem \cite{CK13}.

In this paper we shall use a generalized form of the classical Hill problem in order to determine the equilibrium points and the associated basins of attraction. The multivariate version of the Newton-Raphson iterative scheme will be used for revealing the basins of convergence on several types of two-dimensional planes.

The present article has the following structure: the most important properties of the dynamical system are presented in Section \ref{sys}. The parametric evolution of the position as well as of the stability of the equilibrium points is investigated in Section \ref{param}. The following Section contains the main numerical results, regarding the evolution of the Newton-Raphson basins of convergence. In Section \ref{bee} we demonstrate how the basin entropy of the configuration $(x,y)$ convergence planes evolves as a function of the perturbation parameter. Our paper ends with Section \ref{conc}, where we emphasize the main conclusions of this work.

\section{Description of the dynamical system}
\label{sys}

The classical Hill problem is derived for the restricted three-body problem when the mass of the secondary body is substantially smaller than that of the primary (e.g., the Sun-Earth system). In the vicinity of the secondary the corresponding potential, of the planar model, is given by
\begin{equation}
V(x,y) = \frac{3}{2}x^2 + \frac{1}{r},
\label{pot0}
\end{equation}
where of course $r = \sqrt{x^2 + y^2}$.

According to \cite{KP17}, Eq. (\ref{pot0}) can be generalized, in a straightforward manner, as follows
\begin{equation}
V(x,y) = \frac{1}{2}\left(x^2 + y^2 \right) + \frac{1}{r} + \epsilon \left(x^2 - \frac{y^2}{2}\right).
\label{pot}
\end{equation}
We observe that when $\epsilon = 1$ the potential (\ref{pot}) is reduced to the classical form of Eq. (\ref{pot0}).

It should be noted that the potential of the generalized Hill problem is composed of very interesting terms. In particular, the first term is an isotropic harmonic oscillator, the second term is a repulsive Coulomb potential, while the third term breaks the rotational symmetry of the oscillator. For $\epsilon$ different from zero, but still smaller than 1, the first and the third parts together form an anisotropic harmonic oscillator. We could say that $\epsilon$ is the perturbation parameter for the rotational symmetry. This is true because for $\epsilon$ approaching 1 from below the system changes its qualitative character and it is no longer a harmonic oscillator with a repulsive Coulomb potential in the middle. On the other hand, for $\epsilon = 1$, or larger, it becomes unbound, for sufficiently large energy levels.

The equations describing the motion of the test particle, in the corotating frame of reference, read
\begin{align}
\ddot{x} &= V_x + 2 \dot{y}, \nonumber\\
\ddot{y} &= V_y - 2 \dot{x},
\label{eqmot}
\end{align}
where
\begin{align}
V_x &= \frac{\partial V}{\partial x} = x\left(1 - \frac{1}{r^3} + 2\epsilon\right), \nonumber\\
V_y &= \frac{\partial V}{\partial y} = y\left(1 - \frac{1}{r^3} -\epsilon\right).
\label{der1}
\end{align}

Similarly, the partial derivatives of the second order, which will be needed later for the multivariate Newton-Raphson iterative scheme, read
\begin{align}
V_{xx} &= \frac{\partial^2 V}{\partial x^2} = 1 - \frac{1}{r^3} + \frac{3x^2}{r^5} + 2\epsilon, \nonumber\\
V_{xy} &= \frac{\partial^2 V}{\partial x \partial y} = \frac{3xy}{r^5}, \nonumber\\
V_{yx} &= \frac{\partial^2 V}{\partial y \partial x} = V_{xy}, \nonumber\\
V_{yy} &= \frac{\partial^2 V}{\partial y^2} = 1 - \frac{1}{r^3} + \frac{3y^2}{r^5} - \epsilon.
\label{der2}
\end{align}

\begin{figure*}[!t]
\centering
\resizebox{\hsize}{!}{\includegraphics{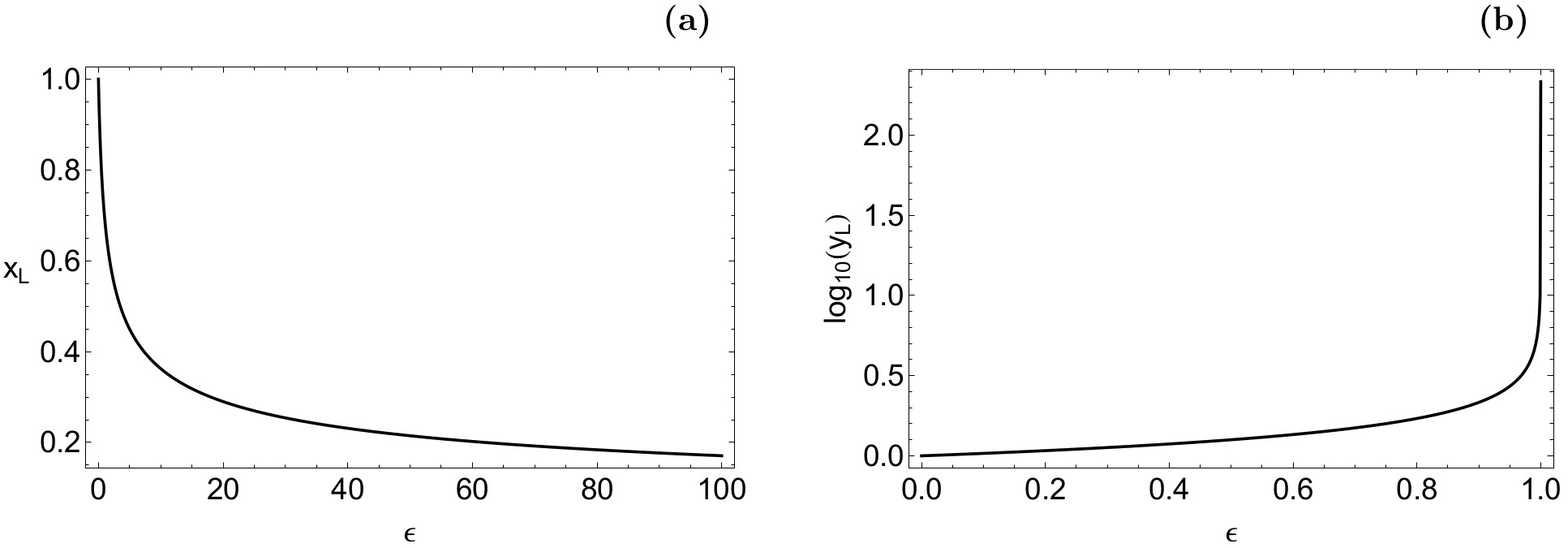}}
\caption{The parametric evolution of (a-left): $x_L$ (b-right): $y_L$, of the equilibrium points in the generalized Hill problem, when $\epsilon \in (0, 100]$. Note that the vertical axis in panel (b) displays the common logarithm of $y_L$.}
\label{evol}
\end{figure*}

The total orbital energy of the system is preserved, according to the Jacobi integral of motion
\begin{equation}
J(x,y,\dot{x},\dot{y}) = 2 V(x,y) - \left(\dot{x}^2 + \dot{y}^2 \right) = \Gamma,
\label{ham}
\end{equation}
where $\dot{x}$ and $\dot{y}$ are the velocities, while $\Gamma$ is the numerical value of the Jacobi constant which is conserved.

\section{Parametric evolution of the equilibrium points}
\label{param}

For the existence of equilibrium points the necessary and sufficient conditions which must be fulfilled are
\begin{equation}
\dot{x} = \dot{y} = \ddot{x} = \ddot{y} = 0.
\label{lps0}
\end{equation}
For determining the coordinates $(x,y)$ of the coplanar equilibrium points we have to numerically solve the following system of equations
\begin{equation}
V_x(x,y) = 0, \ \ \ V_y(x,y) = 0.
\label{lps}
\end{equation}

The total number of the equilibrium points in the generalized Hill problem in not constant but it strongly depends on the value of the perturbation parameter $\epsilon$. More precisely
\begin{itemize}
  \item When $\epsilon \in (0, 1)$ there are four equilibrium points. Two of them, $L_1$ and $L_2$, are located on the $x$-axis, while the other two, $L_3$ and $L_4$, are located on the vertical $y$-axis.
  \item When $\epsilon \geq 1$ there are only two real equilibrium points on the horizontal $x$-axis.
\end{itemize}
In both cases the equilibrium points $L_1$ and $L_2$ are located at $(\pm x_L,0)$, while the coordinates of $L_3$ and $L_4$ are $(0, \pm y_L)$.

It would be very interesting to obtain the exact evolution of the coordinates of the libration points as a function of the perturbation parameter $\epsilon$, when $\epsilon > 0$. Our numerical analysis is illustrated in Fig. \ref{evol}(a-b), where the parametric evolution of $x_L$ and $y_L$ is given as a function of $\epsilon$. It is seen that when $\epsilon \to 0$ both $x_L$ and $y_L$ tend to 1. However, as the value of the perturbation increases the coordinates follow a different path. More precisely, $x_L$ is reduced and tends asymptotically to zero, when $\epsilon \to \infty$. On the other hand, $y_L$ tends asymptotically to infinity, when $\epsilon \to 1$.

In order to determine the linear stability of an equilibrium point the origin of the reference frame must be transferred at the exact position $(x_0,y_0)$ of the libration point through the transformation
\begin{equation}
x = x_0 + \xi, \ \ \ y = y_0 + \eta.
\label{trans}
\end{equation}
The next step is to expand the system of the equations of motion (\ref{eqmot}) into first-order terms, with respect to $\xi$ and $\eta$.
\begin{equation}
\dot{{\bf{\Xi}}} = A {\bf{\Xi}}, \ \ {\bf{\Xi}} = \left(\xi, \eta, \dot{\xi}, \dot{\eta}\right)^{\rm T},
\label{ls}
\end{equation}
where ${\bf{\Xi}}$ is the state vector of the test particle with respect to the equilibrium points, while $A$ is the time-independent coefficient matrix of variations
\begin{equation}
A =
\begin{bmatrix}
    0 & 0 & 1 & 0 \\
    0 & 0 & 0 & 1 \\
    V_{xx}^0 & V_{xy}^0 &  0 & 2 \\
    V_{yx}^0 & V_{yy}^0 & -2 & 0
\end{bmatrix},
\end{equation}
where the superscript 0, at the partial derivatives of second order, denotes evaluation at the position of the equilibrium point $(x_0, y_0)$. The new linearized system describes infinitesimal motions near an equilibrium point.

The characteristic equation of the linear system (\ref{ls}) is
\begin{equation}
\alpha \lambda^4 + b \lambda^2 + c = 0,
\label{ceq}
\end{equation}
where
\begin{align}
\alpha &= 1, \nonumber\\ 
b &= 4 - V_{xx}^0 - V_{yy}^0, \nonumber \\ 
c &= V_{xx}^0 V_{yy}^0 - V_{xy}^0 V_{yx}^0.
\end{align}
It is seen that equation (\ref{ceq}) is quadratic with respect to $\Lambda = \lambda^2$ and therefore it can be written as
\begin{equation}
\alpha \Lambda^2 + b \Lambda + c = 0.
\label{ceq2}
\end{equation}

The necessary and sufficient condition for an equilibrium point to be linearly stable is all four roots of the characteristic equation (\ref{ceq}) to be pure imaginary\footnote{In Hamiltonian (symplectic) dynamics the monodromy matrix is symplectic and this restricts the eigenvalues strongly. Then the product of all eigenvalues must be equal to 1. In addition: if $\lambda$ is an eigenvalue, then also - $\lambda$ and the complex conjugate of $\lambda$ and the complex conjugate of - $\lambda$ must be eigenvalues. For the eigenplanes there are the following possibilities: 2-dimensional hyperbolic planes (with two real eigenvalues $\lambda$ and - $\lambda$), 2-dimensional elliptic planes (with two imaginary eigenvalues $\lambda$ and - $\lambda$) and 4-dimensional planes of complex spiralling behaviour or complex instability (with four eigenvalues with the properties as mentioned above). Here it should be noted that asymptotic stability, in the sense of Lyapunov, is not possible for Hamiltonian systems. Therefore, usually elliptic behaviour of Hamiltonian systems is considered stable. However, in the more general sense this stability is not asymptotic stability. It is only marginal or neutral stability, which means that generally speaking trajectories do not disappear exponentially. More details, regarding the linear stability in Hamiltonian systems, can be found in chapter 3 of \cite{AM87}.}. This means that the following three conditions must be simultaneously fulfilled
\begin{equation}
b > 0, \ \ \ c > 0, \ \ \ D = b^2 - 4 a c > 0.
\end{equation}
This fact ensures that equation (\ref{ceq2}) has two real negative roots $\Lambda_{1,2}$, which consequently implies that there are four pure imaginary roots for $\lambda$.

Since we already know the exact positions $(x_0,y_0)$ of the libration points, we can insert them into the characteristic equation (\ref{ceq}) and therefore determine the linear stability of the equilibrium points, through the nature of the four roots. In the interval $\epsilon \in (0,10^{5}]$ we defined a uniform sequence of $10^6$ values of the perturbation parameter $\epsilon$. Then for these values of $\epsilon$ we numerically solved the system (\ref{lps}) thus computing the coordinates $(x_0,y_0)$ of the equilibrium points. The last step was to insert the coordinates of the equilibria into the characteristic equation (\ref{ceq}) and determine the nature of the four roots. The above-mentioned numerical analysis suggests that for all the equilibrium points the characteristic equation (\ref{ceq}) has always, at least, two complex roots with non zero (positive or negative) real part. Therefore we conclude that all the equilibrium points, $L_i$, $i=1,...,4$, are linearly unstable, when $\epsilon > 0$.

\section{The basins of attraction}
\label{bas}

Over the years, many methods for solving numerically systems of non-linear equations have been developed. Perhaps the most well-known method of all is the Newton-Raphson method. A system of multivariate functions $f({\bf{x}}) = 0$ can be solved using the following iterative scheme
\begin{equation}
{\bf{x}}_{n+1} = {\bf{x}}_{n} - J^{-1}f({\bf{x}}_{n}),
\label{sch}
\end{equation}
where $f({\bf{x_n}})$ is the system of equations, while $J^{-1}$ is the corresponding inverse Jacobian matrix. In our case the system of equations is described in Eqs. (\ref{lps}).

The iterative formulae for each coordinate $(x,y)$, derived from scheme (\ref{sch}), are
\begin{align}
x_{n+1} &= x_n - \left( \frac{V_x V_{yy} - V_y V_{xy}}{V_{yy} V_{xx} - V^2_{xy}} \right)_{(x_n,y_n)}, \nonumber\\
y_{n+1} &= y_n + \left( \frac{V_x V_{yx} - V_y V_{xx}}{V_{yy} V_{xx} - V^2_{xy}} \right)_{(x_n,y_n)},
\label{nrm}
\end{align}
where $x_n$, $y_n$ are the values of the $x$ and $y$ coordinates at the $n$-th step of the iterative process.

The numerical algorithm of the Newton-Raphson method works as follows: The code is activated when an initial condition $(x_0,y_0)$ on the configuration plane is inserted, while the iterative procedure continues until an attractor of the system is reached, with the desired accuracy. If the iterative procedure leads to one of the attractors then we say that the method converges for the particular initial condition. However, in general terms, not all initial conditions converge to an attractor of the system. All the initial conditions that lead to a specific final state (attractor) compose the Newton-Raphson basins of attraction, which are also known as basins of convergence or even as attracting regions/domains. At this point, it should be highly noticed that the Newton-Raphson basins of attraction should not be mistaken, by no means, with the classical basins of attraction which exist in the case of dissipative systems. The Newton-Raphson basins of attraction are just a numerical artifact produced by an iterative scheme, while on the other hand the basins of attraction in dissipative systems correspond to a real observed phenomenon (attraction).

Nevertheless, the determination of the Newton-Raphson basins of attraction is very important because they reflect some of the most intrinsic qualitative properties of the dynamical system. This is true because the iterative formulae of Eqs. (\ref{nrm}) contain both the first and second order derivatives of the effective potential function $V(x,y)$.

\begin{figure*}[!t]
\centering
\resizebox{0.80\hsize}{!}{\includegraphics{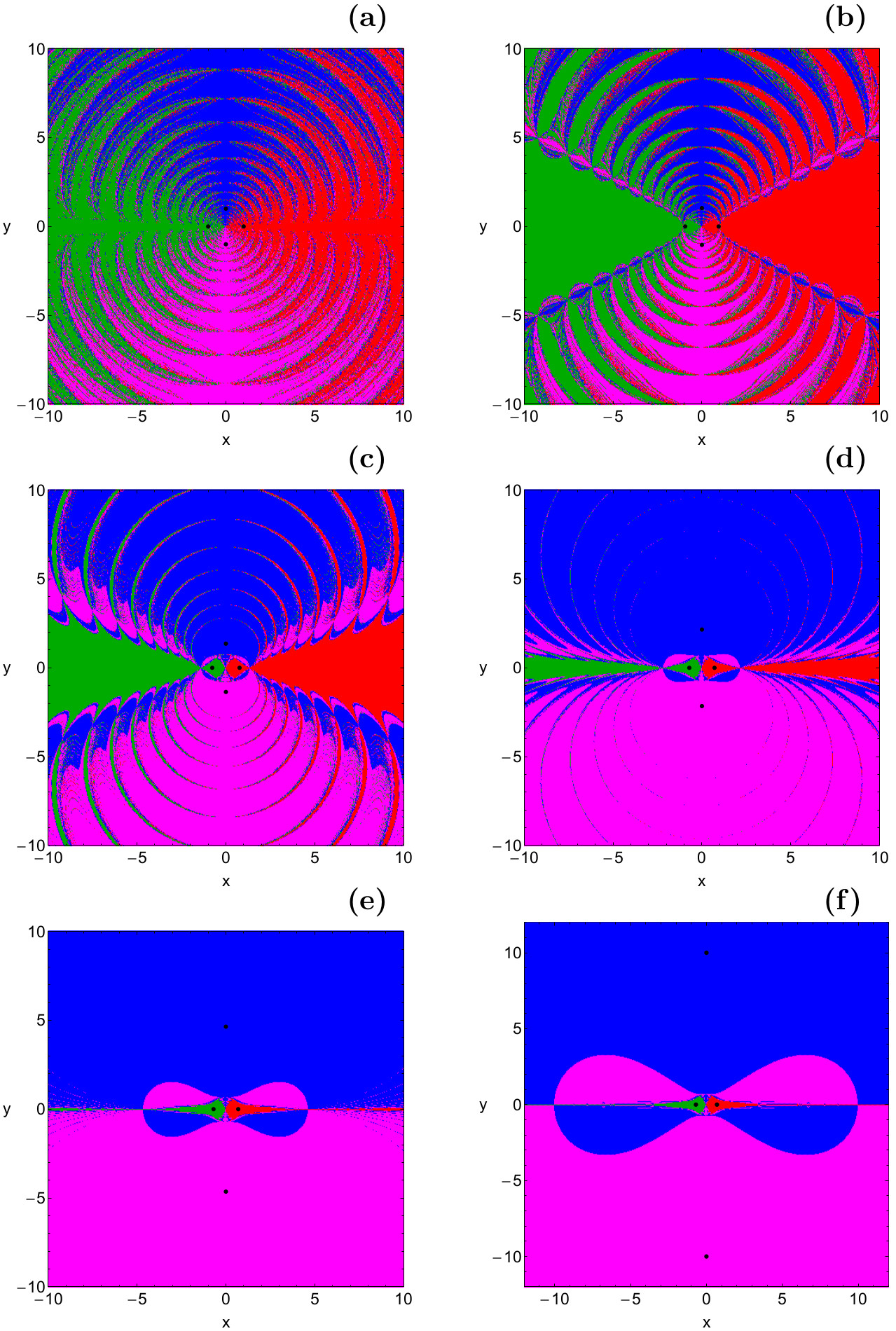}}
\caption{The Newton-Raphson basins of attraction on the configuration $(x,y)$ plane when four equilibrium points are present. (a): $\epsilon = 0.001$; (b): $\epsilon = 0.1$; (c): $\epsilon = 0.6$; (d): $\epsilon = 0.9$; (e): $\epsilon = 0.99$; (f): $\epsilon = 0.999$. The positions of the four equilibrium points are indicated by black dots. The color code, denoting the four attractors is as follows: $L_1$ (green); $L_2$ (red); $L_3$ (blue); $L_4$ (magenta); non-converging points (white).}
\label{c1}
\end{figure*}

\begin{figure*}[!t]
\centering
\resizebox{0.90\hsize}{!}{\includegraphics{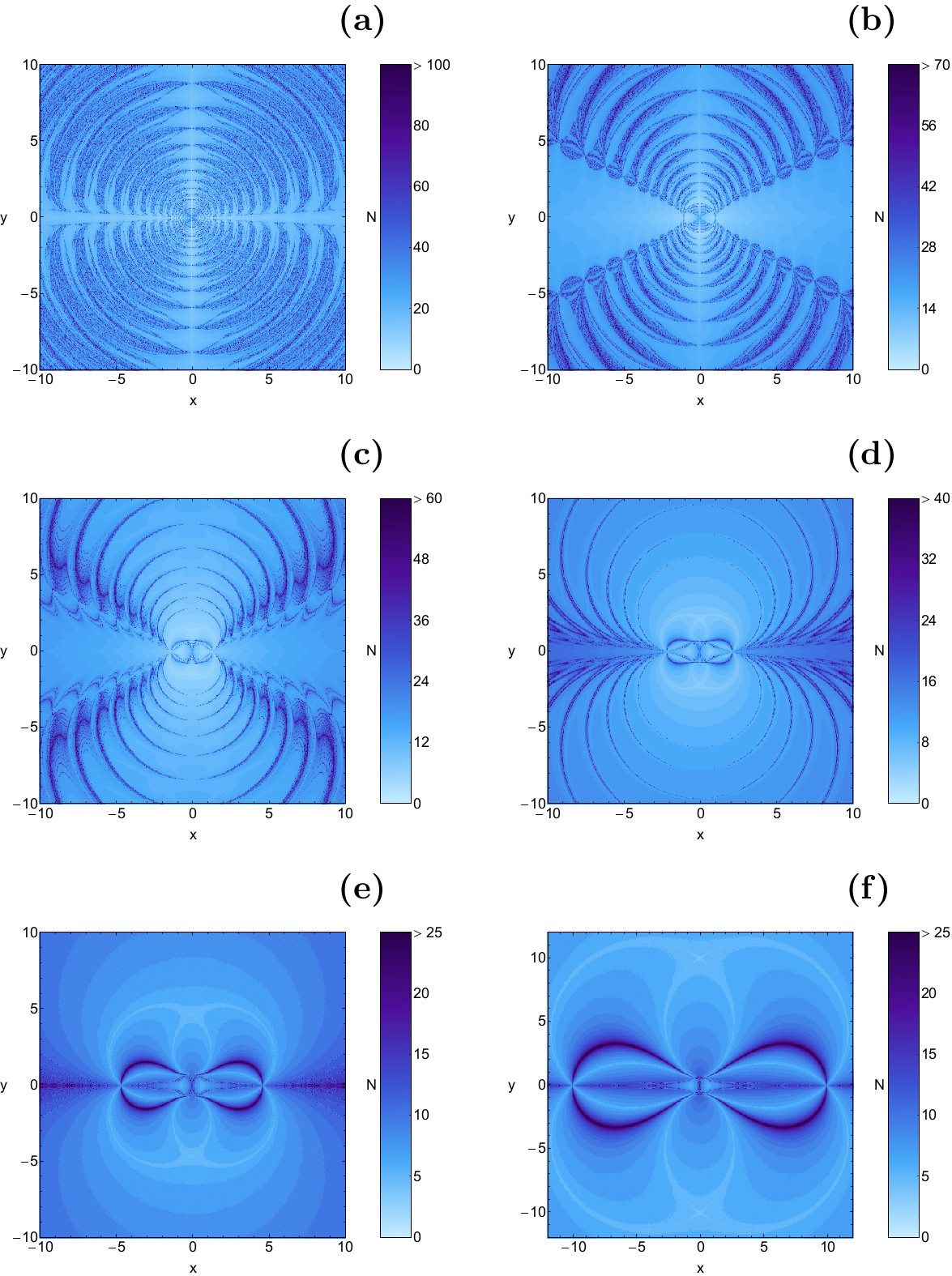}}
\caption{The distribution of the corresponding number $N$ of required iterations for obtaining the Newton-Raphson basins of attraction shown in Fig. \ref{c1}(a-f). The non-converging points are shown in white.}
\label{n1}
\end{figure*}

\begin{figure*}[!t]
\centering
\resizebox{0.80\hsize}{!}{\includegraphics{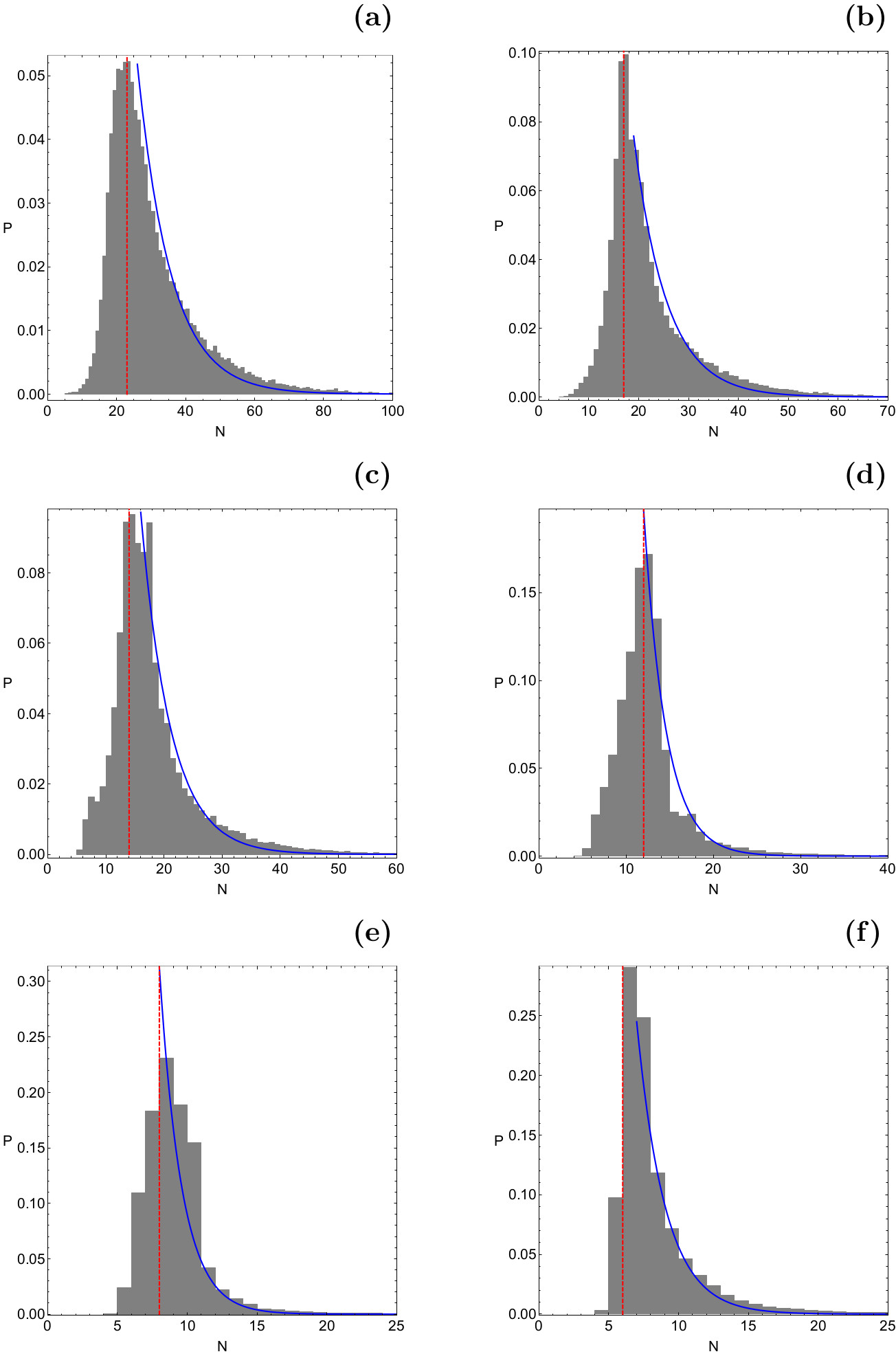}}
\caption{The corresponding probability distribution of required iterations for obtaining the Newton-Raphson basins of attraction shown in Fig. \ref{c1}(a-f). The vertical dashed red line indicates, in each case, the most probable number $N^{*}$ of iterations. The blue line is the best fit for the right-hand side $(N > N^{*})$ of the histograms, using a Laplace probability distribution function.}
\label{p1}
\end{figure*}

In order to unveil the basins of convergence we have to perform a double scan of the configuration $(x,y)$ plane. For this purpose we define uniform grids of $1024 \times 1024$ $(x_0,y_0)$ nodes which shall be used as initial conditions of the numerical algorithm. Of course the initial condition $(0,0)$ is excluded from all grids, because for this initial condition the distance $r$ is equal to zero and consequently several terms, entering formulae (\ref{nrm}), become singular. During the classification of the initial conditions we also keep records of the number $N$ of iterations, required for the desired accuracy. Obviously, the better the desired accuracy, the higher the required iterations. In our calculations the maximum number of iterations is set to $N_{\rm max} = 500$, while the iterative procedure stops only when an accuracy of $10^{-15}$ is reached, regarding the position of the attractors.

In what follows we will try to determine how the perturbation parameter $\epsilon$ influences the structure of the Newton-Raphson basins of attraction in the generalized Hill problem, by considering two cases regarding the total number of the equilibrium points (attractors). For classifying the initial conditions on the configuration $(x,y)$ plane we will use color-coded diagrams (CCDs), where each pixel is assigned a color, according to the final state (attractor) of the initial condition.

\begin{figure*}[!t]
\centering
\resizebox{0.80\hsize}{!}{\includegraphics{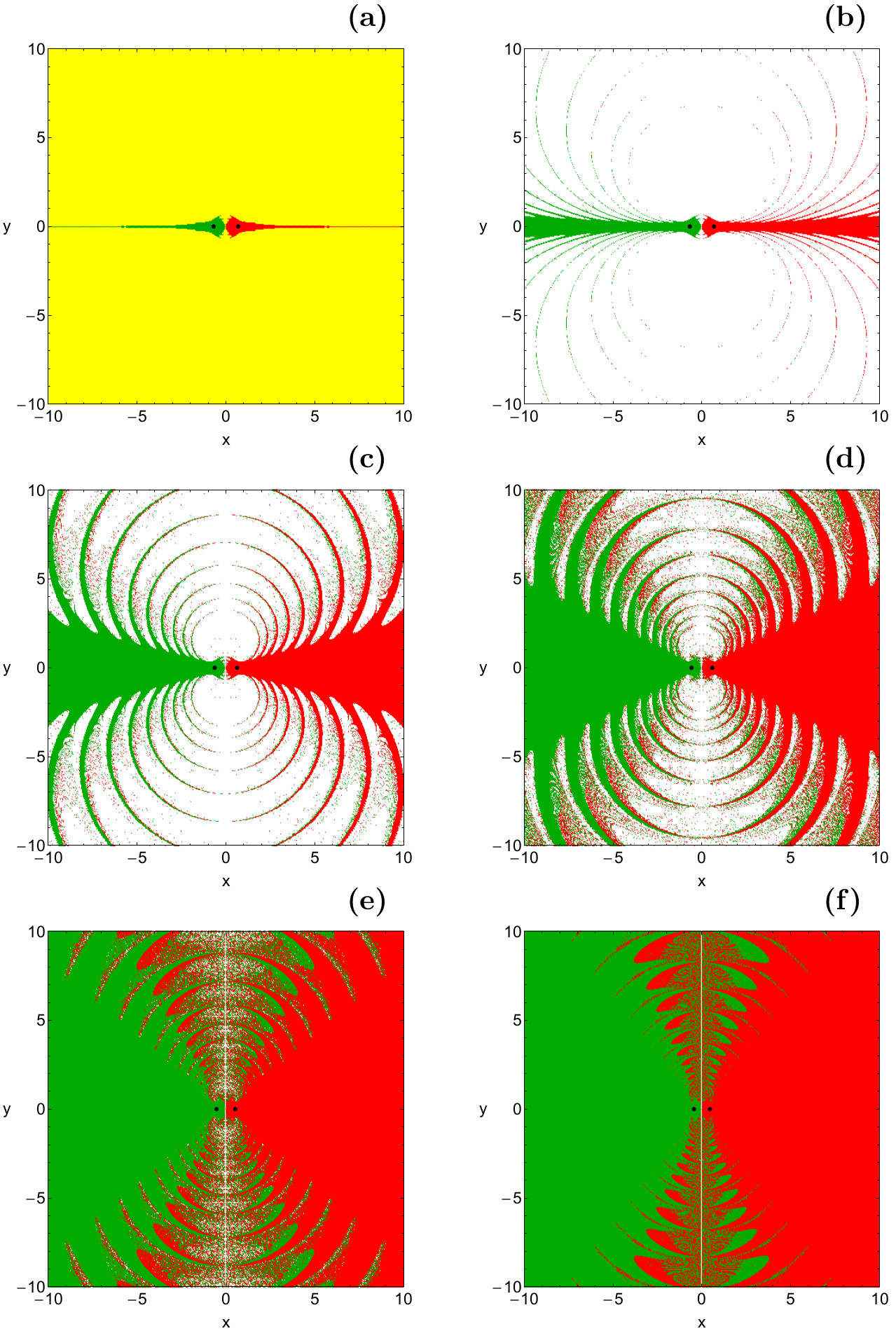}}
\caption{The Newton-Raphson basins of attraction on the configuration $(x,y)$ plane when only two equilibrium points are present. (a): $\epsilon = 1.0$; (b): $\epsilon = 1.1$; (c): $\epsilon = 1.5$; (d): $\epsilon = 2.0$; (e): $\epsilon = 3.0$; (f): $\epsilon = 5.0$. The positions of the two equilibrium points are indicated by black dots. The color code, denoting the four attractors is as follows: $L_1$ (green); $L_2$ (red); converging points to infinity (yellow); non-converging points (white).}
\label{c2}
\end{figure*}

\begin{figure*}[!t]
\centering
\resizebox{0.90\hsize}{!}{\includegraphics{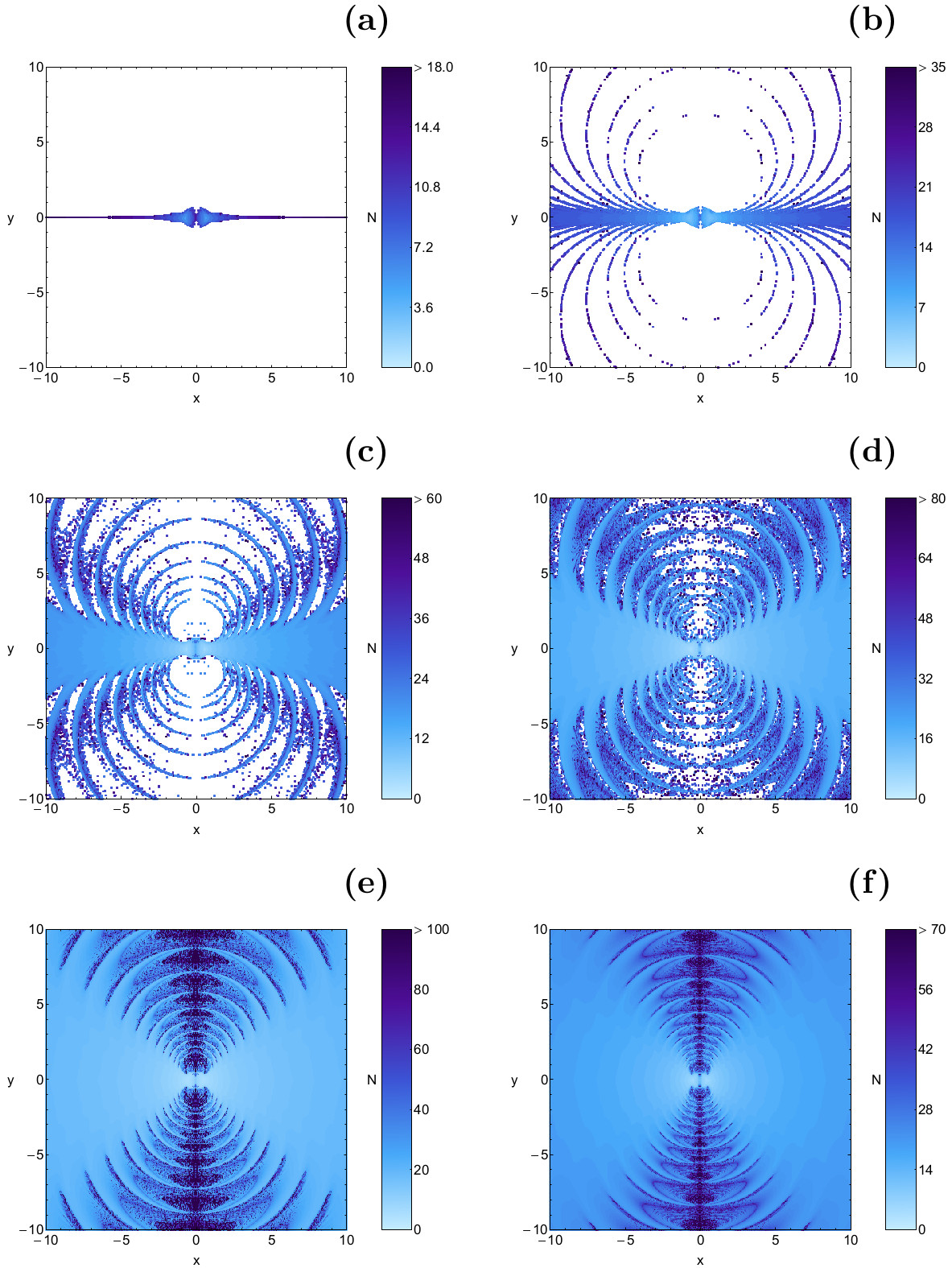}}
\caption{The distribution of the corresponding number $N$ of required iterations for obtaining the Newton-Raphson basins of attraction shown in Fig. \ref{c2}(a-f). The non-converging points, as well as those which converge to infinity, are shown in white.}
\label{n2}
\end{figure*}

\begin{figure*}[!t]
\centering
\resizebox{0.80\hsize}{!}{\includegraphics{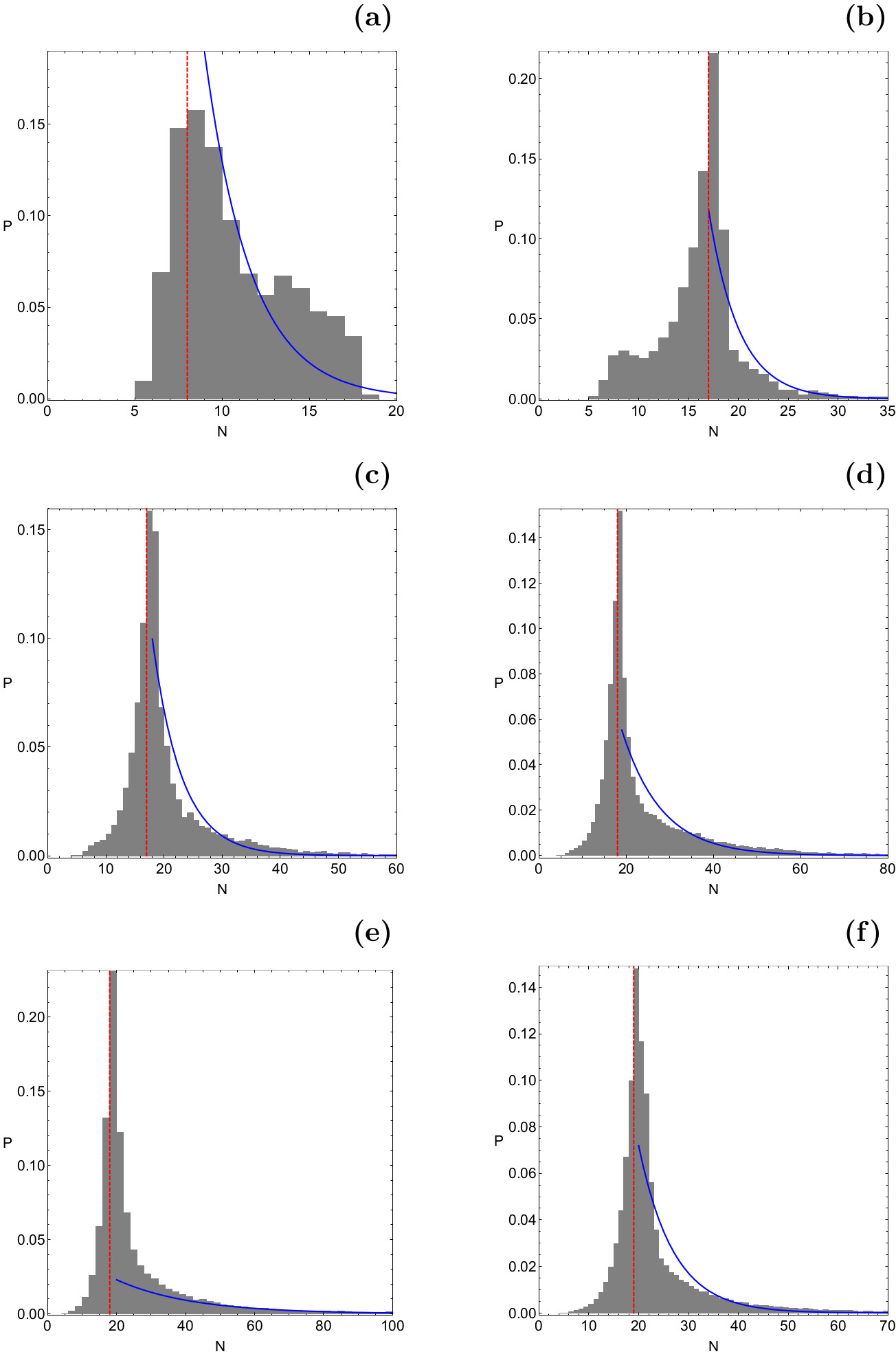}}
\caption{The corresponding probability distribution of required iterations for obtaining the Newton-Raphson basins of attraction shown in Fig. \ref{c2}(a-f). The vertical dashed red line indicates, in each case, the most probable number $N^{*}$ of iterations. The blue line is the best fit for the right-hand side $(N > N^{*})$ of the histograms, using a Laplace probability distribution function.}
\label{p2}
\end{figure*}

\subsection{Case I: Four equilibrium points}
\label{ss1}

Our numerical exploration begins with the first case where four equilibrium points are present, that is when $0 < \epsilon < 1$. In Fig. \ref{c1} we present the evolution of the basins of convergence for six values of the perturbation parameter $\epsilon$. In panel (a) where $\epsilon = 0.001$ we observe the existence of several tentacles between the different basins of attraction. In the vicinity of these tentacles we encounter the most highly fractal areas of the configuration $(x,y)$ plane. At this point it should be noted that when we claim that a region is fractal we simply mean that it has a fractal-like geometry, without conducting, at least for now, any additional quantitative calculations as in \cite{AVS01}. As we proceed to higher values of the perturbation parameter three important phenomena take place
\begin{enumerate}
  \item The area on the configuration $(x,y)$ plane, occupied by the tentacle-like structures, is reduced.
  \item Well formed basins of convergence emerge. In particular, the attracting domains corresponding to the libration points $L_3$ and $L_4$ seem to dominate, while the basins of convergence corresponding to equilibrium points $L_1$ and $L_2$ are mainly confined near the horizontal axis.
  \item The area of the fractal regions on the configuration space is also reduced, thus increasing the predictability regarding the final state (attractor) of the initial conditions.
\end{enumerate}
When $\epsilon$ tends to 1, it is seen in panel (f) that almost the entire $(x,y)$ plane is covered by basins of attraction corresponding to libration points $L_3$ and $L_4$. On the other hand, the attracting regions associated with the equilibrium points $L_1$ and $L_2$ are confined mainly in the vicinity of the corresponding libration points.

The distribution of the corresponding number $N$ of iterations is provided, using tones of blue, in Fig. \ref{n1}(a-f). It is observed that initial conditions inside the attracting regions converge relatively fast, while the slowest converging points are those in the vicinity of the basin boundaries. In particular, the slowest converging points are encountered in the boundaries of either the tentacles or the figure-eight structures observed when $\epsilon > 0.9$. In Fig. \ref{p1}(a-f) the corresponding probability distribution of iterations is given. The probability $P$ is defined as follows: if $N_0$ initial conditions $(x_0,y_0)$ converge to one of the attractors, after $N$ iterations, then $P = N_0/N_t$, where $N_t$ is the total number of initial conditions in every CCD. With increasing value of $\epsilon$ the most probable number $N^{*}$ of iterations is reduced from 23 when $\epsilon = 0.001$ to 6 when $\epsilon = 0.999$. The blue lines in the histograms of Fig. \ref{p1} indicate the best fit to the right-hand side $N > N^{*}$ of them (more details are given in subsection \ref{geno}).

\subsection{Case II: Two equilibrium points}
\label{ss2}

When $\epsilon \geq 1$ there are only two real equilibrium points located on the horizontal $x$-axis. The Newton-Raphson basins of attraction for six values of the perturbation parameter are presented in Fig. \ref{c2}(a-f). A very interesting behavior is unveiled in panel (a), where $\epsilon = 1$. One may observe that the vast majority of the configuration $(x,y)$ plane is covered by initial conditions which converge to extremely large numbers, thus indicating convergence to infinity. This phenomenon however is anticipated. This is true because according to panel (b) of Fig. \ref{evol} when $\epsilon = 1$, $y_L$ tends to infinity. On this basis, we may say that what we see in panel (a) of Fig. \ref{c2} is just a numerical confirmation of the theory.

As soon as $\epsilon > 1$ the convergence properties of the $(x,y)$ plane change drastically. More precisely:
\begin{itemize}
  \item A portion of the configuration space is covered by initial conditions which do not converge to any of the two equilibrium points. Additional computations suggest that for these initial conditions the multivariate Newton-Raphson scheme does not display any sign of convergence even after a substantial amount of iterations $(N = 10000)$.
  \item With increasing value of the perturbation parameter the fractal regions of the $(x,y)$ plane are heavily been reduced, while the areas where the unpredictability is still high are mainly confined near the vertical $y$-axis, around the non-converging initial conditions.
\end{itemize}

\begin{figure}[!t]
\centering
\resizebox{\hsize}{!}{\includegraphics{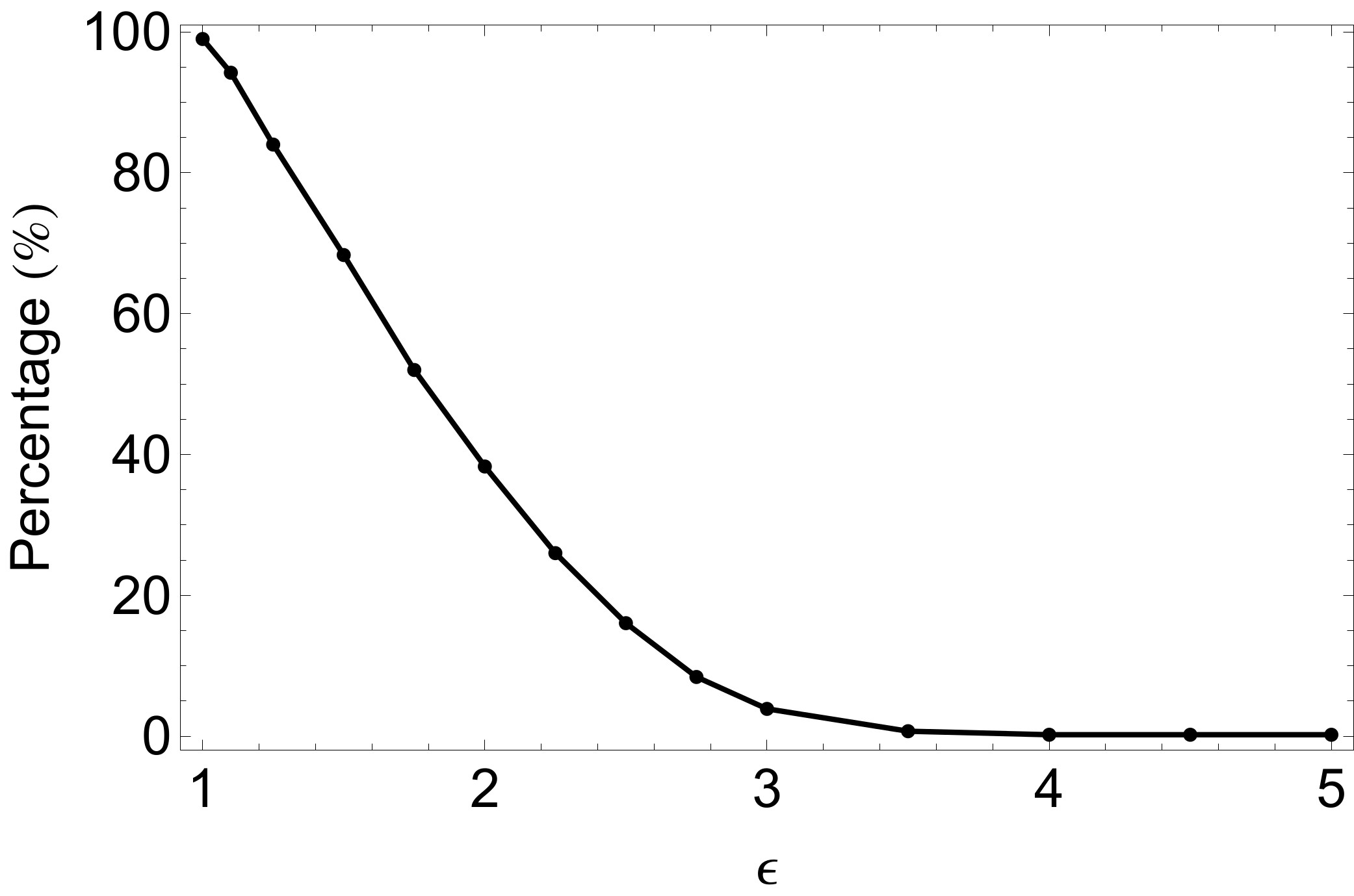}}
\caption{Evolution of the percentage of non-converging initial conditions, as a function of the perturbation parameter $\epsilon$, when $\epsilon \in (1,5]$.}
\label{nfc}
\end{figure}

\begin{figure}[!t]
\centering
\resizebox{\hsize}{!}{\includegraphics{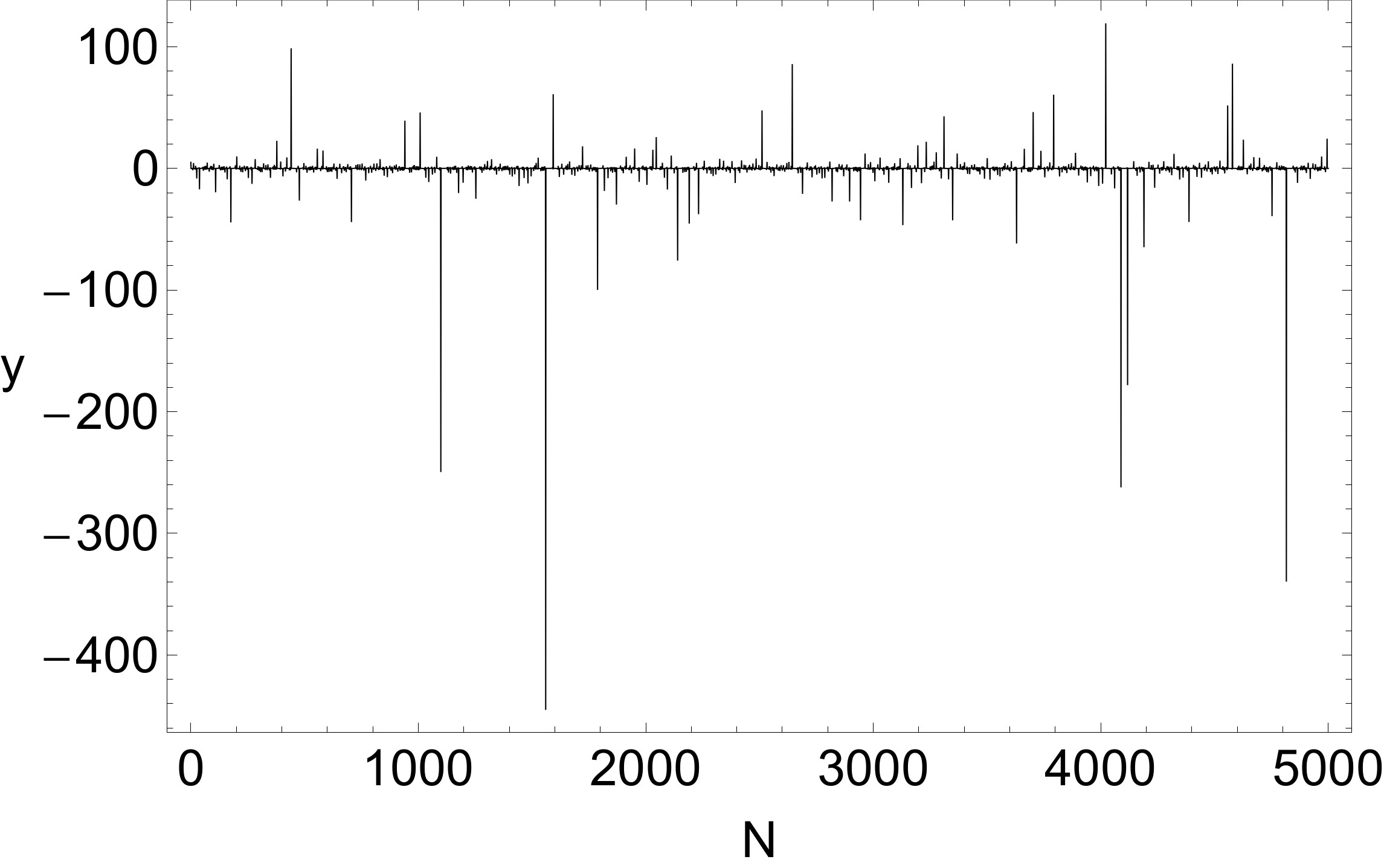}}
\caption{Evolution of the vertical coordinate $y$, as a function of the iterations $N$, when $x_0 = 0$, $y_0 = 5$, and $\epsilon = 5$. Note the chaotic fluctuations as well as the random peaks throughout the range of iterations which clearly indicate non-convergence of the Newton-Raphson iterative scheme.}
\label{nc}
\end{figure}

In Fig. \ref{n2}(a-f) we illustrate the distribution of the corresponding number $N$ of iterations required for obtaining the desired accuracy, while the corresponding probability distribution of iterations is given in Fig. \ref{p2}(a-f). In this case, the most probable number $N^{*}$ of iteration starts at 8 for $\epsilon = 1$ and then it increases up to 19 when $\epsilon = 5$.

Looking at the panels of Fig. \ref{c2} it becomes evident that the amount of non-converging points is reduced with increasing value of the perturbation parameter. In Fig. \ref{nfc} we provide the evolution of the percentage of the non-converging initial conditions as a function of $\epsilon$. We clearly see that the reduction is very smooth and almost linear for $0 < \epsilon < 3$. For larger values of the perturbation parameter $(\epsilon > 4)$ the portion of the non-converging initial conditions remains constant at about 0.22\%, while for $\epsilon > 5$ all initial conditions which do not converge to any of the attractors (equilibrium points) lie completely on the vertical $y$-axis, with $x = 0$.

In \cite{Z17a}, where we investigated the basins of attraction in the planar equilateral restricted four-body problem, we encountered the phenomenon of slow convergence, that is when for an initial condition the Newton-Raphson iterative scheme requires an extremely high number of iterations in order to converge to one of the attractors. Additional numerical computations strongly suggest that this is not the case in the generalized Hill problem. To prove this we set the maximum allowed number of iterations equal to 5000 and we repeated the classification of the initial conditions. However we found that the percentage of non-converging initial conditions remains the same. In Fig. \ref{nc} we illustrate a characteristic example of the evolution of the $y$ coordinate of an non-converging initial condition with $x_0 = 0$ and $y_0 = 5$, when $\epsilon = 5$. It is seen that the vertical coordinate randomly fluctuates between negative and positive numbers, while displaying completely random peaks thus showing no indication of convergence. The same pattern appears even after 50000 iterations which automatically leads to the conclusion that the non-converging initial conditions that are present when $\epsilon > 1$ are true non-converging points.

\subsection{An overview analysis}
\label{geno}

\begin{figure*}[!t]
\centering
\resizebox{\hsize}{!}{\includegraphics{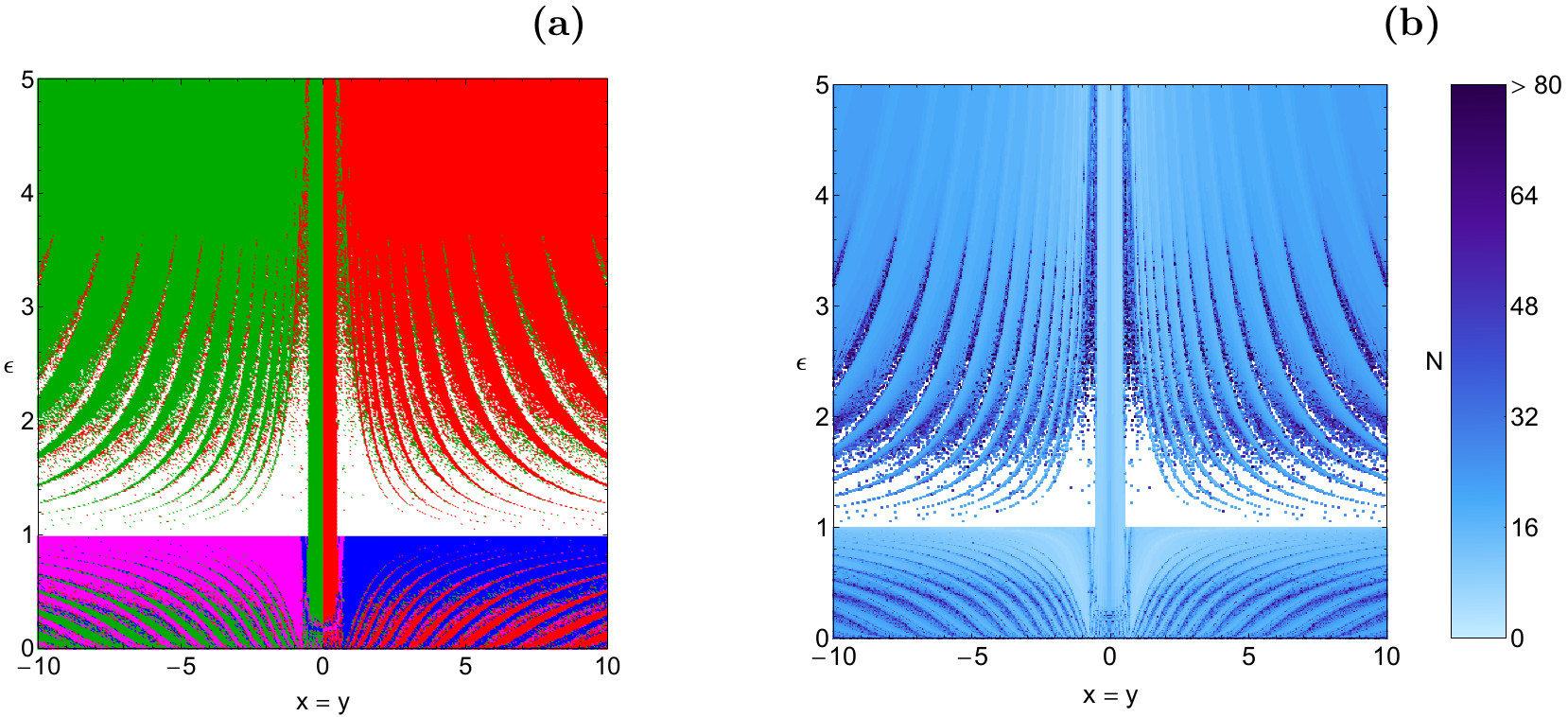}}
\caption{(a-left): The Newton-Raphson basins of attraction on the $(x = y,\epsilon)$ plane, when $\epsilon \in (0,5]$. The color code denoting the attractors is the same as in Fig. \ref{c1}. (b-right): The distribution of the corresponding number $N$ of required iterations for obtaining the basins of convergence shown in panel (a).}
\label{xye}
\end{figure*}

The color-coded convergence diagrams on the configuration $(x,y)$ space, presented earlier in Figs. \ref{c1} and \ref{c2} provide sufficient information regarding the attracting domains however for only a fixed value of the perturbation parameter $\epsilon$. In order to overcome this handicap we can define a new type of distribution of initial conditions which will allow us to scan a continuous spectrum of $\epsilon$ values rather than few discrete levels. The most interesting configuration is to set $x = y$, while the value of the perturbation parameter will vary in the interval $(0,5]$. This technique allows us to construct, once more, a two-dimensional plane in which the $x$ or the $y$ coordinate is the abscissa, while the value of $\epsilon$ is always the ordinate. Panel (a) of Fig. \ref{xye} shows the attracting domains of the $(x = y,\epsilon)$ plane, while in panel (b) of the same figure the distribution of the corresponding number $N$ of required iterations for obtaining the Newton-Raphson basins of attraction is shown. In panel (a) of Fig. \ref{xye} it can be seen very clearly how the convergence properties of the system change when $\epsilon = 1$. At the same time we observe how the tentacles emerge and divide the several basins of attraction.

Additional interesting information could be extracted from the probability distributions of iterations presented in Figs. \ref{p1}, and \ref{p2}. In particular, it would be very interesting to try to obtain the best fit of the tails\footnote{By the term ``tails" of the distributions we refer to the right-hand side of the histograms, that is, for $N > N^{*}$.} of the distributions. For fitting the tails of the histograms, we used the Laplace distribution, which is the most natural choice, since this type of distribution is very common in systems displaying transient chaos (e.g., \cite{ML01,SS08,SASL06}). Our calculations strongly indicate that in the vast majority of the cases the Laplace distribution is the best fit to our data. The only two cases where the Laplace distribution fails to properly fit the corresponding numerical data are the cases corresponding to $\epsilon = 1$ and $\epsilon = 3$ (see panels (a) and (e) of Fig. \ref{p2}, respectively).

The probability density function (PDF) of the Laplace distribution is given by
\begin{equation}
P(N | a,b) = \frac{1}{2b}
 \begin{cases}
      \exp\left(- \frac{a - N}{b} \right), & \text{if } N < a \\
      \exp\left(- \frac{N - a}{b} \right), & \text{if } N \geq a
 \end{cases},
\label{pdf}
\end{equation}
where $a$ is the location parameter, while $b > 0$, is the diversity. In our case we are interested only for the $x \geq a$ part of the distribution function.

In Table \ref{t1} we present the values of the location parameter $a$ and the diversity $b$, as they have been obtained through the best fit, for all cases discussed in Figs. \ref{p1}, and \ref{p2}. One may observe that for most of the cases the location parameter $a$ is very close to the most probable number $N^{*}$ of iterations, while in some cases these two quantities coincide.

\begin{table}[!ht]
\begin{center}
   \caption{The values of the location parameter $a$ and the diversity $b$, related to the most probable number $N^{*}$ of iterations, for all the studied cases shown in Figs. \ref{p1}, and \ref{p2}.}
   \label{t1}
   \setlength{\tabcolsep}{10pt}
   \begin{tabular}{@{}lrrrr}
      \hline
      Figure & $\epsilon$ & $N^{*}$ & $a$ & $b$ \\
      \hline
      \ref{p1}a & 0.0001 & 23 & $N^{*} + 3$ & 9.65 \\
      \ref{p1}b &    0.1 & 17 & $N^{*} + 2$ & 6.59 \\
      \ref{p1}c &    0.6 & 14 & $N^{*} + 2$ & 5.14 \\
      \ref{p1}d &    0.9 & 12 & $N^{*}    $ & 2.54 \\
      \ref{p1}e &   0.99 &  8 & $N^{*}    $ & 1.59 \\
      \ref{p1}f &  0.999 &  6 & $N^{*} + 1$ & 2.04 \\
      \hline
      \ref{p2}a & 1.0 &  8 & $N^{*} + 1$ &  2.64 \\
      \ref{p2}b & 1.1 & 17 & $N^{*} - 1$ &  3.03 \\
      \ref{p2}c & 1.5 & 17 & $N^{*} + 1$ &  5.01 \\
      \ref{p2}d & 2.0 & 18 & $N^{*} + 1$ &  9.04 \\
      \ref{p2}e & 3.0 & 18 & $N^{*} + 2$ & 21.81 \\
      \ref{p2}f & 5.0 & 19 & $N^{*} + 1$ &  6.95 \\
      \hline
   \end{tabular}
\end{center}
\end{table}

\section{Parametric evolution of the basin entropy}
\label{bee}

In the previous Section we discussed the fractality of the convergence diagrams using only qualitative arguments. However it would be very informative if we could have quantitative results regarding the evolution of the fractality. In a recent paper \cite{DWGGS16} a new tool for measuring the uncertainty of the basins has been introduced. This new tool is called the ``basin entropy" and refers to the topology of the basins, thus describing the notion of fractality and unpredictability in the context of basins of attraction or basins of escape.

Let us briefly recall the numerical algorithm of the basin entropy. We assume that there are $N(A)$ attractors (equilibrium points) in a certain region $R = [-10,10] \times [-10,10]$ of the configuration space in our dynamical system. Moreover, $R$ can be subdivided into a grid composed of $N$ square boxes. Each box of the square grid can contain between 1 and $N(A)$ attractors. Therefore we can denote $P_{i,j}$ the probability that inside the box $i$ the resulting attractor is $j$. Due to the fact that inside the box the initial conditions are completely independent, the Gibbs entropy, of every box $i$, is given by
\begin{equation}
S_{i} = \sum_{j=1}^{m_{i}}P_{i,j}\log_{10}\left(\frac{1}{P_{i,j}}\right),
\end{equation}
where $m_{i} \in [1,N_{A}]$ is the number of the attractors inside the box $i$.

The entropy of the entire region $R$, on the configuration $(x,y)$ space, can be computed as the sum of the entropies of the resulting $N$ boxes of the square grid as $S = \sum_{i=1}^{N} S_{i}$. On this basis, the entropy relative to the total number of boxes $N$, which is called basin entropy $S_{b}$, is given explicitly by the following expression
\begin{equation}
S_{b} = \frac{1}{N}\sum_{i=1}^{N}\sum_{j=1}^{m_{i}}P_{i,j}\log_{10}\left(\frac{1}{P_{i,j}}\right).
\end{equation}

\begin{figure}[!t]
\centering
\resizebox{\hsize}{!}{\includegraphics{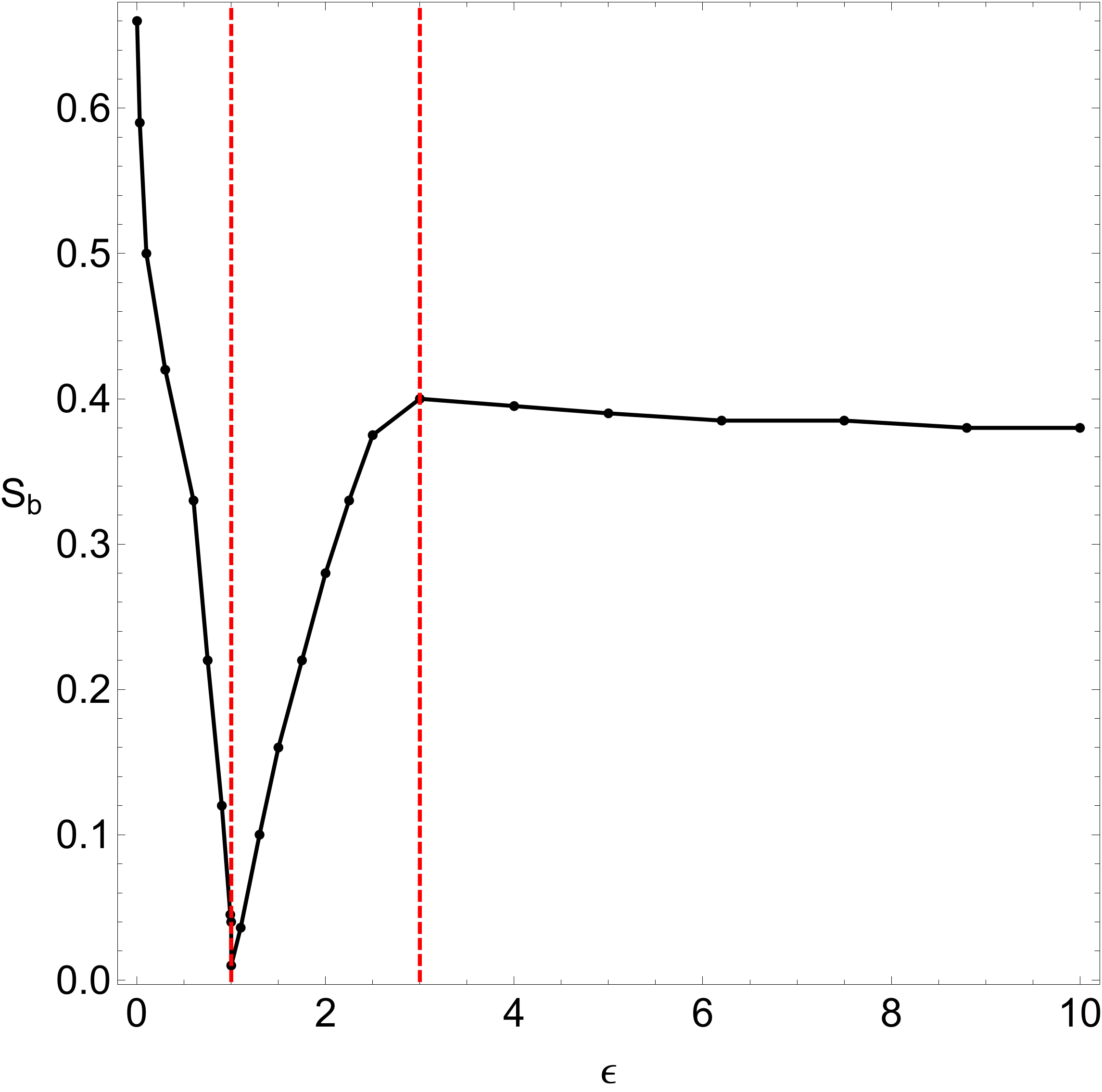}}
\caption{Evolution of the basin entropy $S_b$, of the configuration $(x,y)$ space, as a function of the perturbation parameter $\epsilon$. The vertical, dashed, red lines delimit the three intervals, regarding the tendency of the parametric evolution of the basin entropy.}
\label{be}
\end{figure}

Using the above-mentioned expressions and also adopting the value $\varepsilon = 0.005$, suggested in \cite{DWGGS16}, we computed the basin entropy $S_{b}$ of the configuration $(x,y)$ plane for several values of the perturbation parameter $\epsilon$. Here it should be clarified that in the case where non-converging points are present, we count them as an additional basin which coexists with the other basins, corresponding to the equilibrium points. In Fig. \ref{be} we present the evolution of the basin entropy as a function of the perturbation parameter. At this point, it should be noted that for creating this diagram we used numerical results not only for the cases presented earlier in Figs. \ref{c1} and \ref{c2} but also from additional values of $\epsilon$. We see that
\begin{itemize}
  \item When $\epsilon \to 0$ the basins get more complicated, which results in an increase of the basin entropy, which displays its maximum value around 0.7. However as the value of the perturbation increases the value of $S_b$ decreases rapidly and when $\epsilon = 1$ the basin entropy is almost zero, since the basins look very smooth (see panel (a) in Fig. \ref{c2}).
  \item When $\epsilon > 1$ new fractal structures emerge and the basin entropy increases almost linearly up to $\epsilon = 4$, where $S_b \approx 0.4$.
  \item When $\epsilon > 4$ it is seen that the value of the basin entropy seems to saturate (reaches a plateau), thus displaying a constant value at around 0.39.
\end{itemize}

\section{Discussion and conclusions}
\label{conc}

The aim of this work was to numerically compute the basins of attraction, associated with the libration points, in the generalized Hill problem. Of paramount importance was the determination of the influence of the perturbation parameter $\epsilon$ on the position as well as on the stability of the equilibrium points. Using the multivariate Newton-Raphson iterative scheme we managed to reveal the beautiful structures of the basins of convergence on several types of two-dimensional planes. The role of the attracting domains is very important since they describe how each initial condition is attracted by the equilibrium points of the system, which act as attractors. Our numerical investigation allowed us to monitor the evolution of the geometry as well as of the fractality of the basins of convergence as a function of the perturbation parameter. Moreover, the basins of attraction have been successfully related with both the corresponding distributions of the number of required iterations, and the probability distributions.

As far as we know, there are no previous studies on the Newton-Raphson basins of convergence in the generalized Hill problem. Therefore, all the presented numerical outcomes of the current thorough and systematic analysis are novel and this is exactly the importance and the contribution of our work.

The most important outcomes of our numerical analysis can be summarized as follows:
\begin{enumerate}
  \item The perturbation parameter strongly influences the dynamical properties of the system. When $0 < \epsilon < 1$ four equilibrium points exist, while for $\epsilon \geq 1$ there are only two real libration points.
  \item Our computations indicate that all the equilibrium points of the system are always unstable.
  \item In all examined cases, regarding the numerical value of the perturbation parameter $\epsilon$, the basins of attraction corresponding to all equilibrium points extend to infinity.
  \item When $\epsilon > 1$ we detected a portion of non-converging initial conditions. Additional numerical calculation (by setting a much higher limit of allowed iterations) revealed that these initial conditions are initial conditions for which the iterative scheme fails to converge to one of the attractors of the system.
  \item The iterative method was found to converge very fast $(0 \leq N < 15)$ for initial conditions around each equilibrium point, fast $(15 \leq N < 25)$ and slow $(25 \leq N < 50)$ for initial conditions that complement the central regions of the very fast convergence, and very slow $(N \geq 50)$ for initial conditions of dispersed points lying either in the vicinity of the basin boundaries, or between the dense regions of the equilibrium points.
  \item As the value of $\epsilon$ increases from 0 to 1 the most probable number of required iterations, $N^{*}$, was found to decrease, while for $\epsilon > 1$ the tendency is reversed.
  \item It was observed that the basin entropy of the configuration $(x,y)$ plane is highly influenced by the perturbation parameter. More precisely, the highest value of $S_b$ is exhibited when $\epsilon \to 0$, while on the other hand the basin entropy tends to zero when $\epsilon \to 1$.
\end{enumerate}

For all the calculation, regarding the determination of the basins of attraction, we used a double precision numerical code, written in standard \verb!FORTRAN 77! \cite{PTVF92}. Furthermore, the latest version 11.2 of Mathematica$^{\circledR}$ \cite{W03} was used for creating all the graphical illustration of the paper. For the classification of each set of initial conditions, in all types of two-dimensional planes, we needed about 5 minutes of CPU time, using an Intel$^{\circledR}$ Quad-Core\textsuperscript{TM} i7 2.4 GHz PC.

We hope that the present numerical outcomes to be useful in the active field of basins of convergence in dynamical systems. Since our present exploration, regarding the attracting domains in the generalized Hill problem, was encouraging it is in our future plans to expand our investigation. In particular, it would be of great interest to try other types of iterative formulae (i.e., of higher order, with respect to the classical iterative method of Newton-Raphson) and determine how they influence the geometry of the basins of convergence. Additionally, the disconnected Wada property (e.g., \cite{KY91,DWSY15}) is a very common feature in Newton-Raphson schemes. Therefore, we could certainly examine if the basins of attraction in the generalized Hill problem have also this striking topological property.

\section*{Acknowledgments}

I would like to express my warmest thanks to the anonymous referees for the careful reading of the manuscript and for all the apt suggestions and comments which allowed us to improve both the quality and the clarity of the paper.


\end{document}